\documentclass[journal,a4paper]{IEEEtran}

\usepackage[cmex10]{amsmath}
\usepackage{amsfonts}
\usepackage{graphicx}
\graphicspath{{../figures-proof-of-concept/}}
\graphicspath{{../figures-asym-rtt/}}
\graphicspath{{../figures/}}
\usepackage[caption=false,font=footnotesize]{subfig}  

\usepackage{tikz}
\usetikzlibrary{positioning}
\usetikzlibrary{automata,arrows}

\usepackage{algorithmic}

\usepackage{cite}
\usepackage{url}
\usepackage{color}
\definecolor{darkgreen}{rgb}{0,0.4,0}
\definecolor{altblue}{rgb}{0,0.2,0.7}

\begin{document}

\title{Improving PIE's performance over high-delay paths}

%
%
%

\author{
  \IEEEauthorblockN{Nicolas Kuhn$^{\dag, \star}$ and David Ros$^{\ddag}$} \\
  \IEEEauthorblockA
  {
    $^{\dag}$Institut Mines-T\'el\'ecom / T\'el\'ecom Bretagne, IRISA, 2 rue de la Ch\^ataigneraie, 35576 Cesson S\'evign\'e, France \\
    $^{\star}$Centre National d'Etudes Spatiales (CNES), 18 Avenue Edouard Belin, 31400 Toulouse, France \\
    $^{\ddag}$Simula Research Laboratory, P.O.\ Box 134, 1325 Lysaker, Norway
  }
}


\maketitle

\begin{abstract}
Bufferbloat is excessive latency due to over-provisioned network buffers. PIE and CoDel are two recently proposed Active Queue Management (AQM) algorithms, designed to tackle bufferbloat by lowering the queuing delay without degrading the bottleneck utilization. PIE uses a proportional integral controller to maintain the average queuing delay at a desired level; however, large Round Trip Times (RTT) result in large spikes in queuing delays, which induce high dropping probability and low utilization. To deal with this problem, we propose Maximum and Average queuing Delay with PIE (MADPIE). Loosely based on the drop policy used by CoDel to keep queuing delay bounded, MADPIE is a simple extension to PIE that adds deterministic packet drops at controlled intervals. By means of simulations, we observe that our proposed change does not affect PIE's performance when RTT $< 100$~ms. The deterministic drops are more dominant when the RTT increases, which results in lower maximum queuing delays and better performance for VoIP traffic and small file downloads, with no major impact on bulk transfers.



\end{abstract}

\begin{IEEEkeywords}
Active queue management; Internet latency; PIE. 
\end{IEEEkeywords}


\section{Introduction}
\label{sec-introduction}

\IEEEPARstart{A}{ctive} queue management (AQM) schemes can be introduced in network routers with the goal of controlling the amount of buffering and reducing the loss synchronization. Large buffers and the absence of AQM deployment have resulted in huge latencies; this problem is known as bufferbloat~\cite{bufferbloat}. 
Since AQM can control the queuing delay, its deployment ``can significantly reduce the latency across an Internet path''~\cite{rfc_recommandations_aqm}. The first AQM proposals, such as Random Early Detection (RED)~\cite{red_algorithm}, dating back more than a decade, have been reported to be usually turned off, mainly because of the difficulty to tune their parameters. Even if Adaptive RED (ARED)~\cite{ared_2001} was proposed to ease the parameterization of RED, it was designed to control the buffering when traffic is composed mainly of TCP flows. Proportional Integral controller Enhanced (PIE)~\cite{pie_algorithm} and Controlled Delay (CoDel)~\cite{codel_algorithm} are two recent AQM schemes that have been designed to tackle bufferbloat by lowering the queuing delay while addressing RED's stability issues and considering the presence of transports that do not react to congestion signals, such as UDP.

PIE and CoDel share two main concepts that can be mapped into two algorithm parameters: (1) a \textit{target delay} ($\tau$) represents the acceptable standing queuing delay above which an AQM  drops packets more aggressively; (2) an update \textit{interval} ($\lambda$) represents the reactivity of an AQM. 
These two parameters have different usages in the two algorithms. In CoDel, $\tau$ embodies an upper bound on the allowed queuing delay; if the minimum queuing delay over an interval of duration $\lambda$ is higher than $\tau$, then a packet is dropped with probability~1, else no packet is dropped. PIE uses $\tau$ to increase or decrease a dropping \emph{probability}, based on the deviations of estimated queuing delay from such target delay: $\tau$ is therefore the desired average queuing delay.

CoDel has been shown to have auto-tuning issues and its performance is sensitive to the traffic  load \cite{jarvinen-lcn-2014}. Also, its default $5$\,ms of maximum allowed queuing delay can be damaging for low-speed bottlenecks~\cite{2015-kuhn-eucnc} and its interval value is based on the assumption that the Round Trip Time (RTT) is $100$\,ms~\cite{ietf-codel}. On the other hand, PIE has been shown to be less sensitive to traffic loads~\cite{jarvinen-lcn-2014}, its default target delay of $20$\,ms should be less problematic with low capacity bottlenecks, and it does not make assumptions on the RTT. However, in this paper, we show that PIE is sensitive to the RTT, as we observe wide oscillations in queuing delay when the RTT increases. This results in temporarily high maximum queuing delay, high dropping probability, and low bottleneck utilization. 

To reduce the RTT sensitivity of PIE and improve the performance of latency sensitive applications over large RTT paths (e.g., rural broadband or satellite access),  
our proposal, Maximum and Average queuing Delay with PIE (MADPIE) extends PIE by adding deterministic drops to prevent the queuing delay from growing beyond a critical value, loosely mimicking CoDel's drop policy.

The rest of this article is organized as follows. Section~\ref{sec-hybrid-aqm} details the MADPIE algorithm. In Section~\ref{sec-proof-of-concept}, by means of simulations we illustrate the issues that PIE faces when the RTT increases, and how the deterministic drops in MADPIE help to correct those issues. Section~\ref{sec-eval} provides an evaluation of the trade-off between allowing more bandwidth for bulk transfers and improving the performance of latency sensitive applications with MADPIE and PIE, as opposed to DropTail. 
Section~\ref{sec-rtt-mix} compares CoDel, PIE and MADPIE when the flows sharing the bottleneck do not have the same RTT. 
Finally, Section~\ref{sec-conclusions} concludes this work. 


\section{Adding deterministic drops to PIE}
\label{sec-hybrid-aqm}

PIE drops an incoming packet when $p \leq p_{drop}$, where $p$ is drawn at random from a uniform distribution in $[0,1]$, and $p_{drop}$ is an internal variable  updated every $\lambda=30$\,ms according to:
\begin{equation}
  \label{eq:pie-drop-proba}
  p_{drop} \leftarrow p_{drop} + \alpha \times (E[T]- \tau) + \beta \times (E[T]- E[T]_{old}).
\end{equation}
$E[T]$ and $E[T]_{old}$ represent the current and previous estimation of the queuing delay. $\tau$ is PIE's target delay. $\alpha$ determines how the deviation of current queuing delay from $\tau$ affects the drop probability, whereas $\beta$ exerts additional adjustments depending on whether the queuing delay is trending up or down. 

MADPIE uses the same random drop policy as PIE, the only difference between the two algorithms being that we add a deterministic drop policy. MADPIE requires only one additional parameter: 
 the queuing delay $\tau_{DD}$ above which deterministic drops occur.
An indicator variable $p_{max}$, initialized to 0, tells whether a packet must be dropped ($p_{max}=1$) or not ($p_{max}=0$) by the deterministic policy. Every $\lambda$, if the estimated queuing delay is $> \tau_{DD}$, $p_{max}$ is set to $1$. Then, if a packet is \emph{not} dropped nor marked by the random drop algorithm and $p_{max}=1$, then a packet is dropped or marked and $p_{max}$ is reset to $0$. Thus, there can be a maximum of one deterministic drop every $\lambda$. 




\section{Proof of concept}
\label{sec-proof-of-concept}

The aim of this section is to illustrate how MADPIE's behaviour differs from that of PIE when the RTT increases.

\begin{figure}[!ht]
\centering
\begin{tikzpicture}[>=stealth',shorten >=1pt,auto,node distance=2.35 cm, scale = 1.25, transform shape]
	\node[scale=0.55, color=blue]         (S) {$snd$};
	\node[state, scale=0.55]         (A) [right of=S,xshift=1cm] {$R_{1}$};
	\node[scale=0.55]         (I) [below of=A, xshift=0.2cm, yshift=1.7cm] {AQM - $Q_{size}=BDP$};
        \node[state, scale=0.55]         (B) [right of=A,xshift=1.5cm] {$R_{2}$};
	\node[scale=0.55, color=blue]         (D) [right of=B,xshift=1cm] {$dest$};
	\path (S)  edge [<->, color=black] node[scale=0.5]{$100$\,Mbps - $1$\,ms} (A);
	\path (A)  edge [ <->, color=red] node[scale=0.5]{$10$\,Mbps - $d$\,ms} (B);
	\path (B)  edge [ <->, color=black] node[scale=0.5]{$100$\,Mbps - $1$\,ms} (D);
\end{tikzpicture}
\caption{Topology used to prove the MADPIE concept.}
\label{fig-topology-proof-of-concept}
\end{figure}
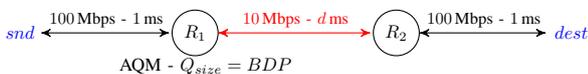

\figurename~\ref{fig-topology-proof-of-concept} presents the topology used in this section. The one-way delay of the bottleneck link is set to $d=48$\,ms or $d=248$\,ms (which corresponds to a base RTT\footnote{That is, the minimum RTT, without any queuing delays.} of $100$\,ms and $500$\,ms, respectively). The queue size at $R_1$ is set to the Bandwidth-Delay Product (BDP). The AQM introduced at $R_1$ is either PIE ($\tau=20$\,ms and $\lambda=30$\,ms) or MADPIE ($\tau=20$\,ms, $\lambda=30$\,ms, $\tau_{DD}=30$\,ms). We simulate 10 TCP bulk flows from $snd$ to $dest$, using CUBIC as congestion control policy, for $300$\,s. The Initial congestion Window (IW) is set to 10~packets and the SACK option is enabled. The flows randomly start in $[0;1]$\,s. All TCP variants used in this article were provided by the NS-2 TCP Linux module updated to linux kernel version \texttt{3.17.4}.\footnote{More details at: \url{http://heim.ifi.uio.no/michawe/research/tools/ns/index.html}}. 

\begin{figure}[!htb]
    \begin{center}
	\subfloat[Queuing delay - PIE. \label{subfig-proof-qdel-cdf-pie}]{\includegraphics[width=0.48\linewidth]{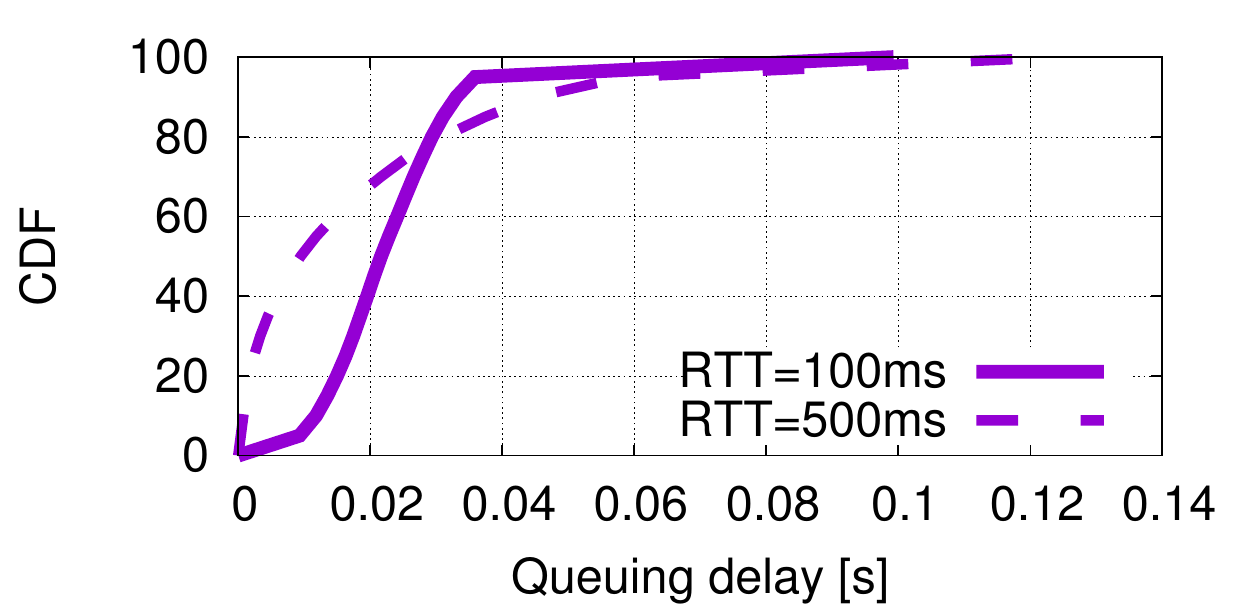}}
	\subfloat[Bottleneck utilization - PIE. \label{subfig-proof-th-cdf-pie}]{\includegraphics[width=0.48\linewidth]{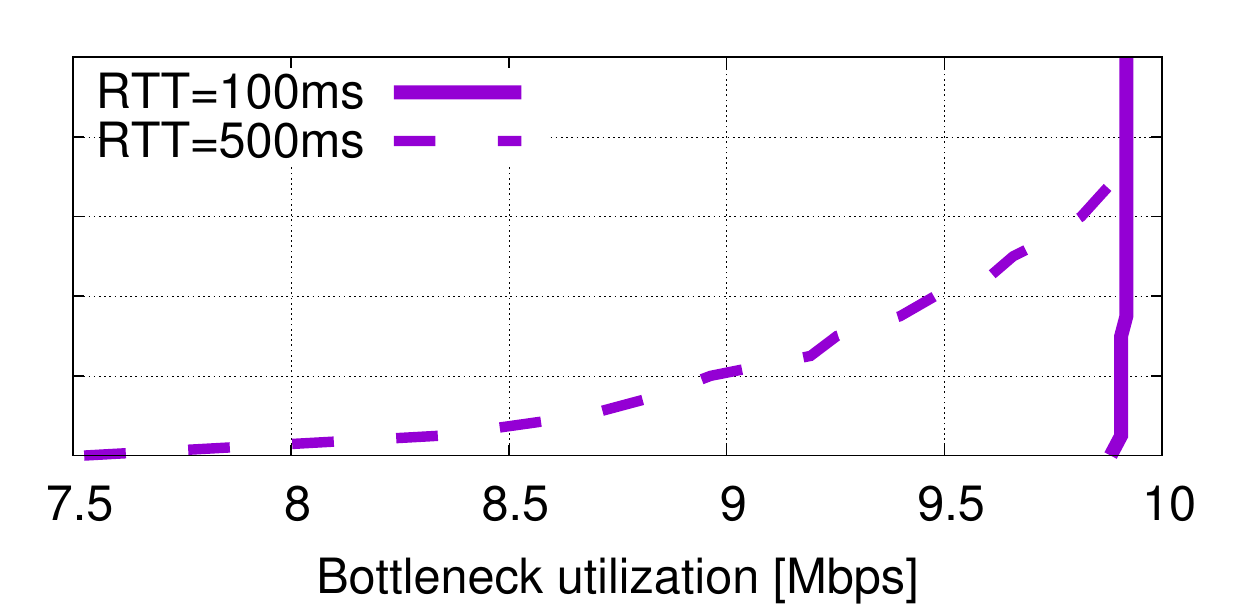}} \\
	\subfloat[Queuing delay - MADPIE. \label{subfig-proof-qdel-cdf-madpie}]{\includegraphics[width=0.48\linewidth]{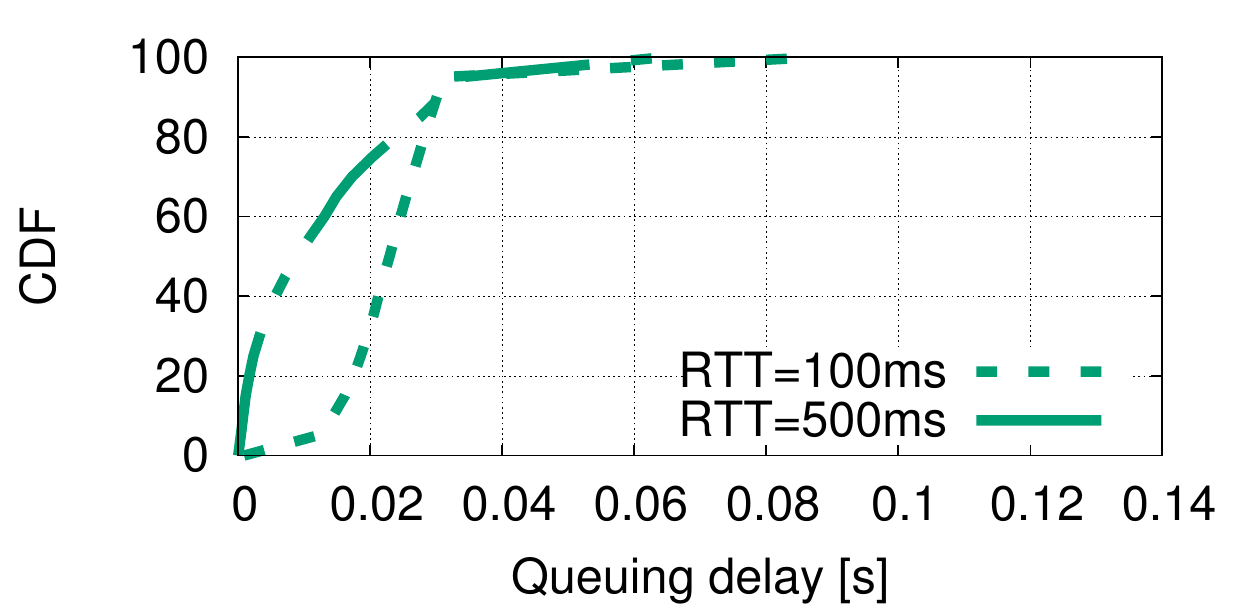}}
	\subfloat[Bottleneck utilization - MADPIE. \label{subfig-proof-th-cdf-madpie}]{\includegraphics[width=0.48\linewidth]{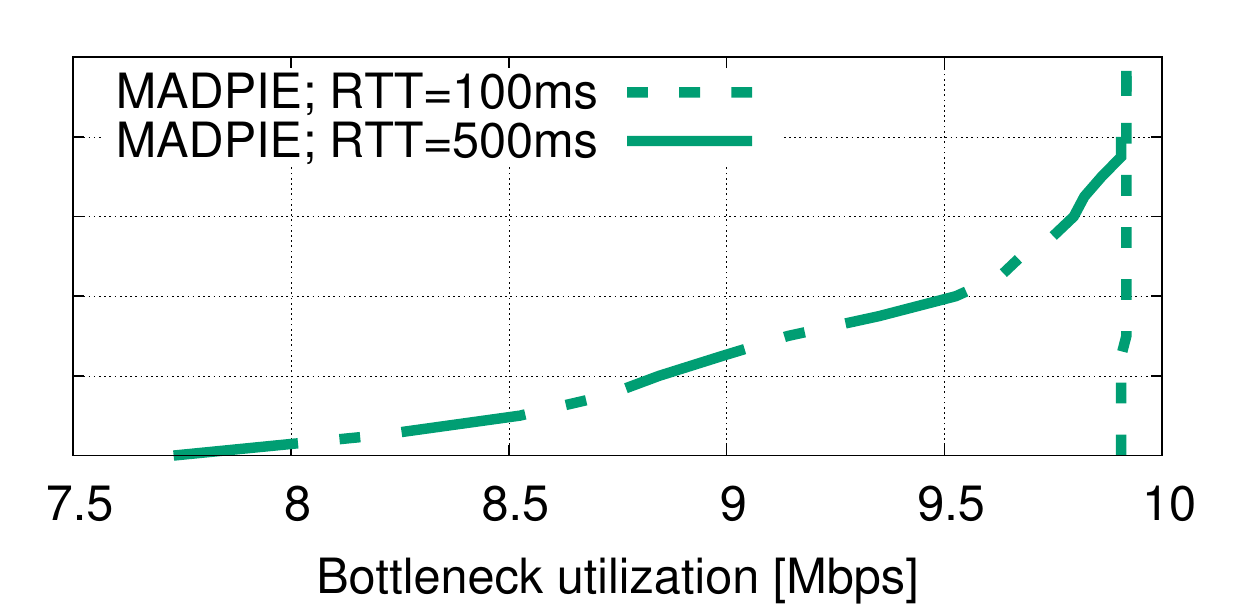}}
        \caption{Bottleneck utilization and queuing delay.}
        \label{fig-proof-bottleneck-cdf}
    \end{center}
\end{figure}


In Fig.~\ref{fig-proof-bottleneck-cdf}, we present the Cumulative Distribution Function (CDF) of the queuing delay (measured per packet) and the bottleneck utilization (sampled every second). When the RTT is $100$\,ms, apart from the maximum queuing delay that is slightly lower with MADPIE than with PIE ($\approx 100$\,ms with PIE, $\approx 80$\,ms with MADPIE), there is no noticeable performance difference between MADPIE and PIE. When the RTT is $500$\,ms, \figurename~\ref{subfig-proof-qdel-cdf-pie} and~\ref{subfig-proof-qdel-cdf-madpie} show that for $20$\,\% of the samples, the queuing delay is higher than $20$\,ms with MADPIE whereas it is higher than $30$\,ms with PIE. Also, with MADPIE as opposed to with PIE, the maximum queuing delay is reduced by $\approx 60$\,ms. It is worth pointing out that $90$\,\% of the samples show a queuing delay lower than $30$\,ms (that is $\tau_{DD}$) with MADPIE as opposed to $50$\,ms with PIE. \figurename~\ref{subfig-proof-th-cdf-pie} and~\ref{subfig-proof-th-cdf-madpie} show that this latency reduction does not induce a lower bottleneck utilization.

\begin{figure}[!htb]
    \begin{center}
	\subfloat[PIE \label{subfig-proof-dyna-pie-500}]{\includegraphics[width=0.48\linewidth]{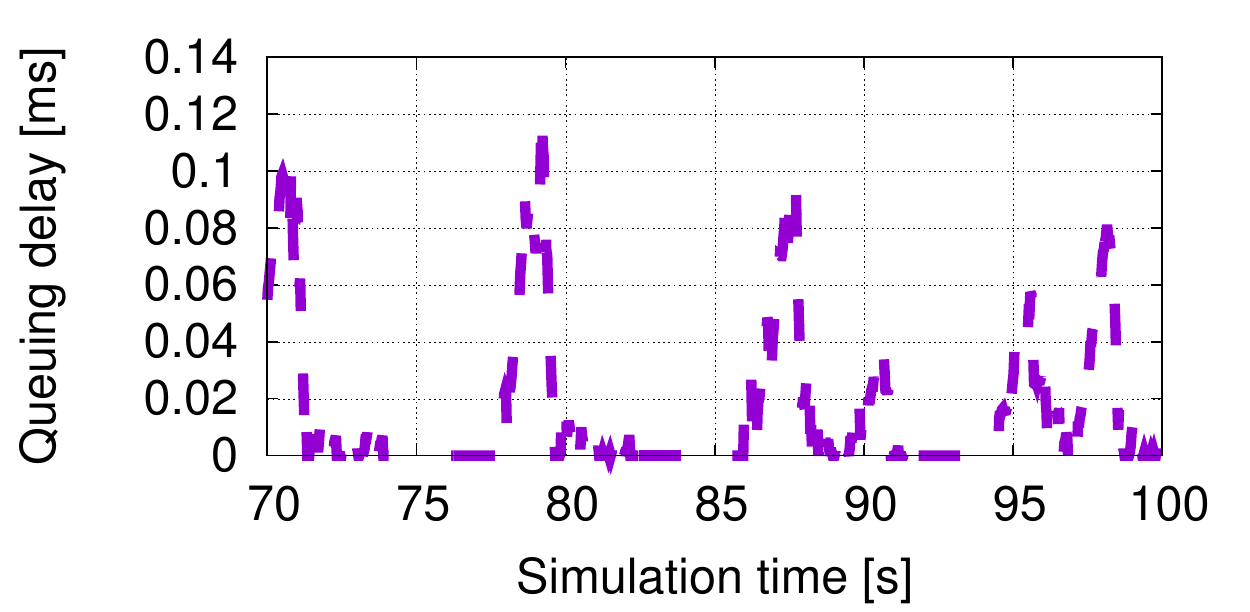}}
	\subfloat[MADPIE \label{subfig-proof-dyna-madpie-500}]{\includegraphics[width=0.48\linewidth]{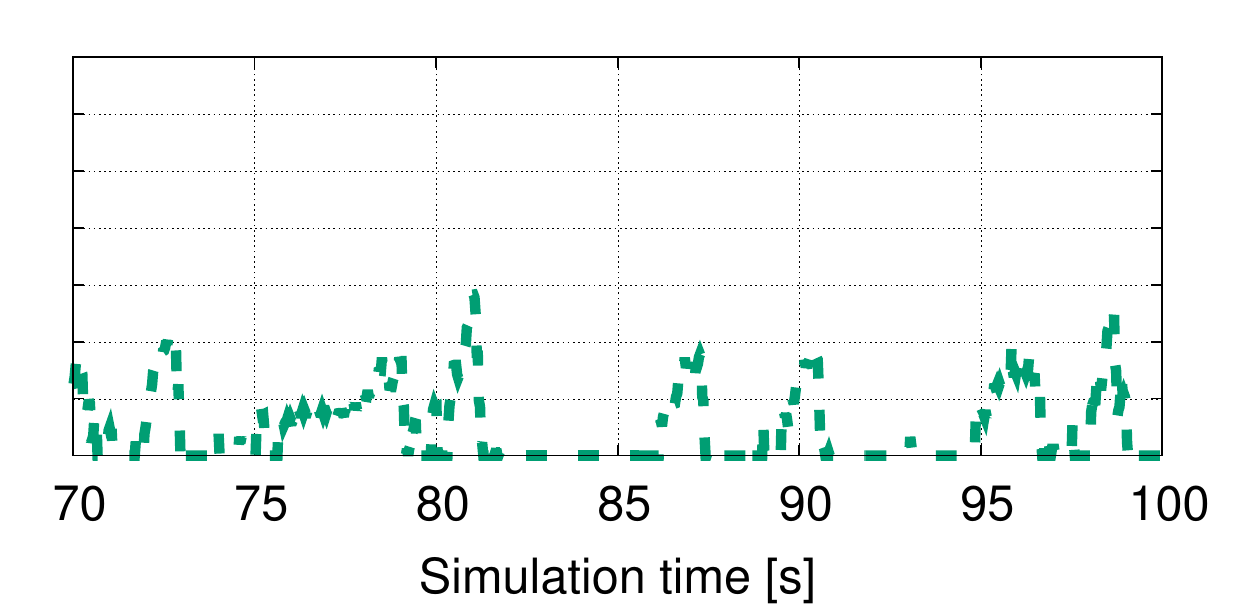}}
        \caption{Evolution of queuing delay over time with $RTT=500$\,ms.}
        \label{fig-proof-qdel-dyna}
    \end{center}
\end{figure}

This latency reduction provided by MADPIE can be further explained by looking at the queuing delay evolution in \figurename~\ref{fig-proof-qdel-dyna}. With PIE, a higher RTT results in wider oscillations in queuing delay: as the queuing delay gets much higher than $\tau$, the dropping probability increases in order to maintain a lower queuing delay. This however results in a momentarily empty buffer. PIE's burst allowance of $100$\,ms lets the queuing delay to frequently grow above $100$\,ms, as the buffer was previously empty. With MADPIE, it is possible to initially allow the same bursts, but the deterministic drops would then prevent an excessive growth of both the queuing delay and the drop probability if the buffer is frequently empty then full, which is what happens when the RTT is $500$\,ms.

\begin{figure}[!htb]
    \begin{center}
	\includegraphics[width=\linewidth]{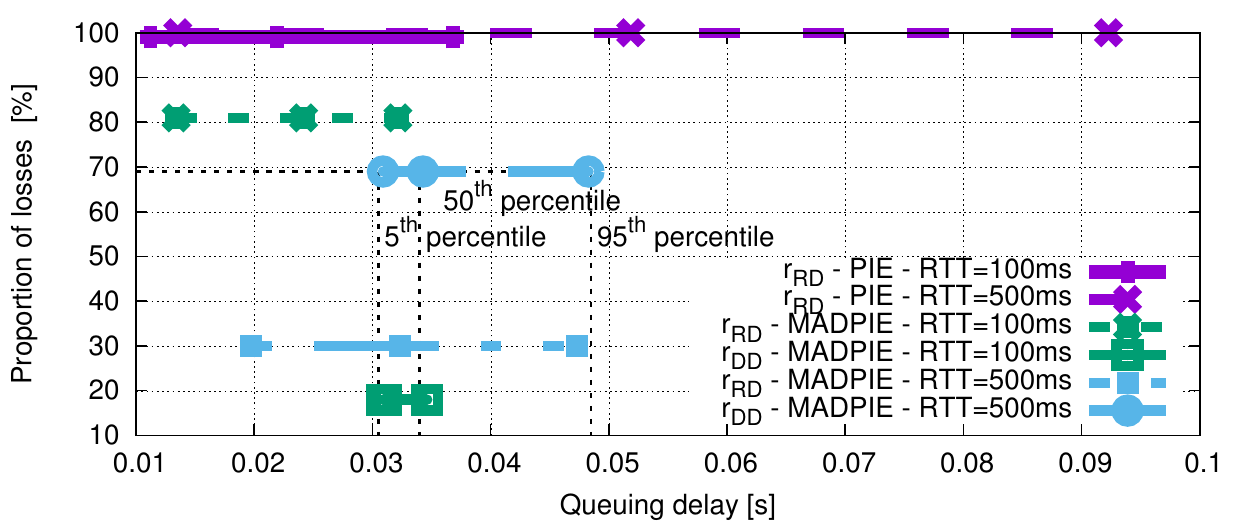}
        \caption{Queuing delay and proportion of losses.}
        \label{fig-proof-qdel-loss}
    \end{center}
\end{figure}

To better understand how MADPIE's behaviour differs from that of PIE when the RTT increases, we
look at the contribution of random and deterministic drops to the overall drop rate. 
Let us denote by $n_{DD}$, $n_{RD}$, $n_{BO}$ and $n_{tot}$ the number of drop events induced by a deterministic drop (only MADPIE), a random drop (PIE and MADPIE), a buffer overflow (PIE and MADPIE) and the total number of drops, respectively. Let $r_{x}=n_{x}/n_{tot}$ be the proportion of drop events of type $x$. \figurename~\ref{fig-proof-qdel-loss} shows $r_{RD}$ and $r_{DD}$ as a function of the queuing delay when the drop occured. As one example (dashed lines in \figurename~\ref{fig-proof-qdel-loss}), when the RTT is $500$\,ms and the AQM is MADPIE, $r_{DD} \approx 70$\,\% and when the deterministic drops occured, the $5$\,\% percentile of the queuing delay was $\approx 30$\,ms, the $50$\,\% percentile $\approx 34$\,ms and the $95$\,\% percentile $\approx 48$\,ms. With PIE, most of the drops are induced by the AQM algorithm and not by buffer overflow and, when the RTT is $500$\,ms, the queuing delay raises up to more than $90$\,ms. With MADPIE, when the RTT is $100$\,ms, the random drop part of MADPIE is responsible for more than $80$\,\% of the drops, whereas when the RTT is $500$\,ms, the deterministic part of MADPIE is responsible for around $70$\,\% of the drops with a consequent queuing delay reduction. 


\section{Performance of MADPIE with a traffic mix}
\label{sec-eval}

We compare now the performance of DropTail (DT), PIE and MADPIE when the traffic comes from a mix of various applications.

\subsection{Traffic and topology}
\label{subsec-eval-traffic-topo}

\begin{figure}[!ht]
\centering
\begin{tikzpicture}[>=stealth',shorten >=1pt,auto,node distance=2.35 cm, scale = 1.25, transform shape]
	\node[scale=0.55, color=blue]         (S) {$snd_{SF}$};
	\node[state, scale=0.55]         (A) [right of=S,xshift=1cm] {$R_{1}$};
	\node[scale=0.55]         (I) [below of=A, xshift=0.5cm, yshift=1.7cm] {AQM - $Q_{size}=BDP$};
        \node[state, scale=0.55]         (B) [right of=A,xshift=1.5cm] {$R_{2}$};
	\node[scale=0.55, color=blue]         (D) [right of=B,xshift=1cm] {$dest_{SF}$};
	\node[scale=0.55, color=magenta]         (E) [above of=D,yshift=-1.4cm] {$dest_{CBR}$};
	\node[scale=0.55, color=green]         (F) [below of=D,yshift=1.4cm] {$dest_{FTP}$};
	\node[scale=0.55, color=magenta]         (G) [above of=S,yshift=-1.4cm] {$snd_{CBR}$};
	\node[scale=0.55, color=green]         (H) [below of=S,yshift=1.4cm] {$snd_{FTP}$};
	\path (S)  edge [<->, color=black] node[scale=0.5]{\small{$100$\,Mbps - $1$\,ms}} (A)
		    (A)  edge [ <->, color=black, sloped, near end, swap] node[scale=0.5]{} (G)
		    (A)  edge [ <->, color=black, near start, sloped, swap ] node[scale=0.5]{} (H)
		    (A)  edge [ <->, color=red] node[scale=0.5]{$10$\,Mbps - $d$\,ms} (B)		    
		    (B)  edge [ <->, color=black] node[scale=0.5]{\small{$100$\,Mbps - $1$\,ms}} (D)
		    (B)  edge [ <->, color=black, sloped, very near end] node[scale=0.5]{} (E)
		    (B)  edge [ <->, color=black, very near start, sloped ] node[scale=0.5]{} (F);		   
\end{tikzpicture}
\caption{Topology and traffic mix used to evaluate MADPIE.}
\label{fig-topology-traff-mix}
\end{figure}
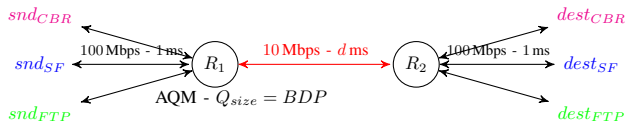

\figurename~\ref{fig-topology-traff-mix} presents the topology used in this section. The one-way delay  $d$ of the bottleneck link is set to $48$\,ms, $148$\,ms or $248$\,ms (\textit{i.e.} a base RTT of $100$\,ms, $300$\,ms or $500$\,ms). The queue size at $R_1$ is set to the BDP. The AQM introduced at $R_1$ is either PIE ($\tau=20$\,ms and $\lambda=30$\,ms) or MADPIE ($\tau=20$\,ms, $\lambda=30$\,ms, $\tau_{DD}=25$\,ms). 

Between $snd_{CBR}$ and $dest_{CBR}$,  there are $N_{CBR}$ Constant Bit-Rate (CBR) UDP flows with a sending rate of $87$\,kbps and a packet size of $218$\,B. The intent is to model Voice-over-IP (VoIP) or gaming traffic, such as in~\cite[p. 17]{cable-labs-traffic}. Between $snd_{SF}$ and $dest_{SF}$, $N_{SF}$ flows transfer files of $S$~kB ($S \in \{ 15; 44; 73; 102 \}$). When a download is finished, a new random value is taken for $S$ and another download starts after $\tau$ seconds, with $\tau$ randomly generated according to an exponential law of mean $9.5$\,s. This traffic lets us assess the benefits of using MADPIE for short flows. Between $snd_{FTP}$ and $dest_{FTP}$, $N_{FTP}$ TCP bulk flows are generated. TCP flows use CUBIC congestion control, and TCP options are the same as those specified in \S~\ref{sec-proof-of-concept}. All the flows randomly start between $0$ and $1$\,s. Each run lasts $100$\,s and is repeated $20$ times with independent seeds. The metrics are sampled every second (except for the queuing delay and the one way delay that are sampled per-packet). We choose to present the results with $N_{CBR}=4$, $N_{SF}=20$ and $N_{FTP}=10$, as this traffic mix stresses both PIE and MADPIE.

\subsection{CBR traffic}
\label{subsec-eval-cbr}

\begin{figure*}[!ht]
    \begin{center}
	\subfloat[Example. \label{fig-traf-mix-cbr-ex}]{\includegraphics[width=0.25\linewidth]{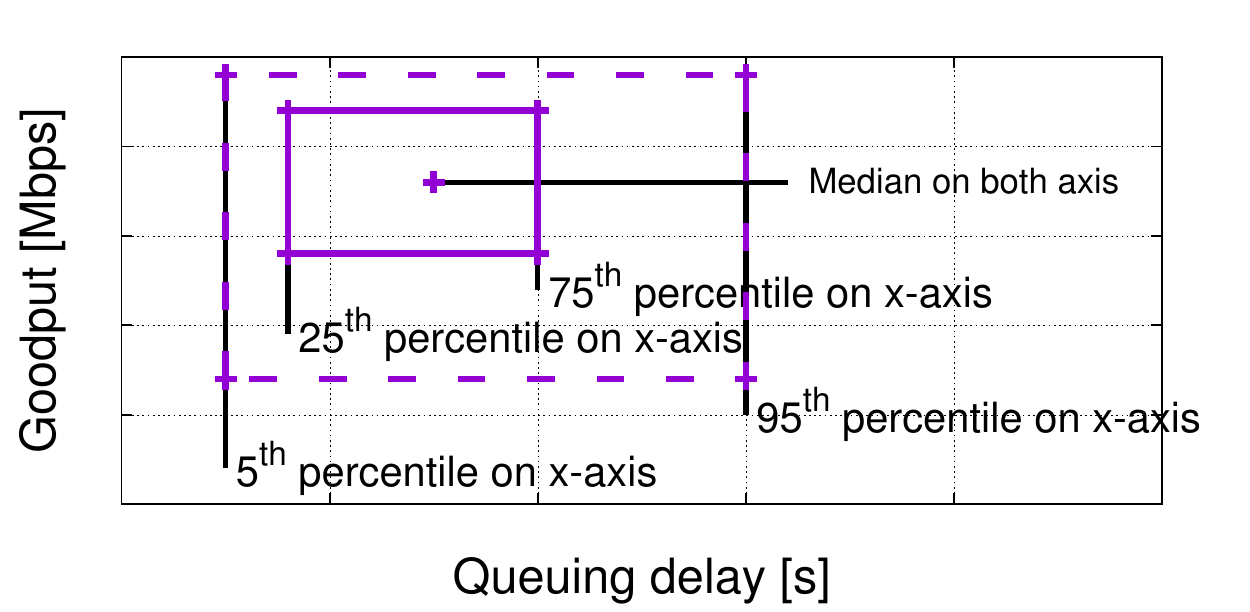}}
	\subfloat[DT. \label{subfig-traf-mix-cbr-droptail}]{\includegraphics[width=0.25\linewidth]{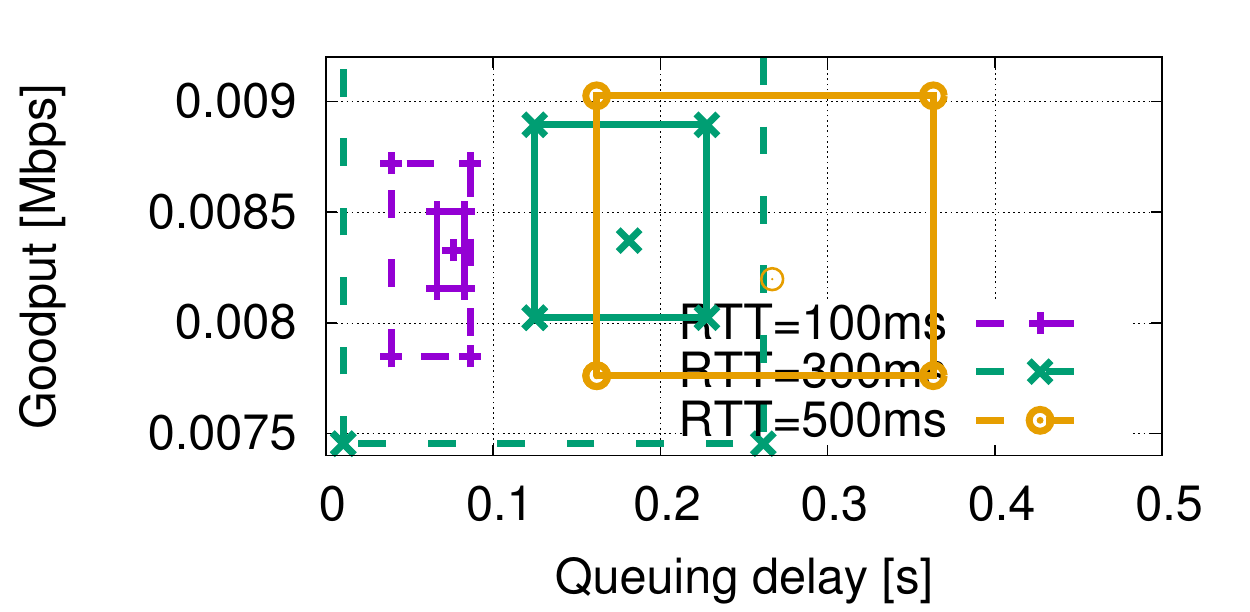}}
	\subfloat[PIE. \label{subfig-traf-mix-cbr-pie}]{\includegraphics[width=0.25\linewidth]{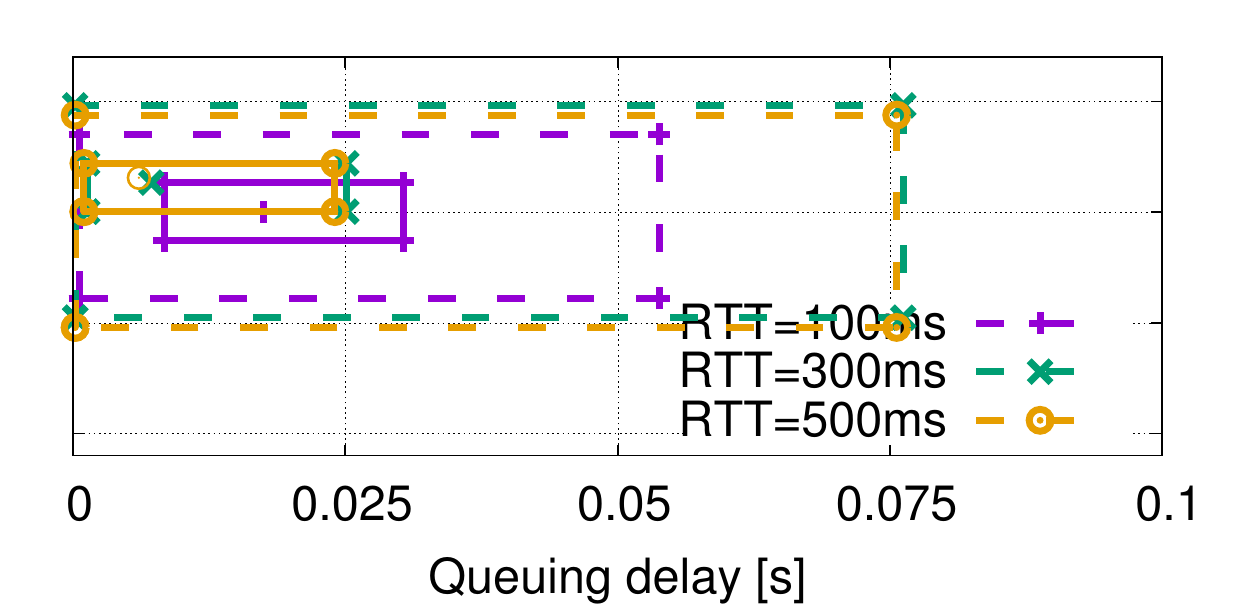}}
	\subfloat[MADPIE. \label{subfig-traf-mix-cbr-madpie}]{\includegraphics[width=0.25\linewidth]{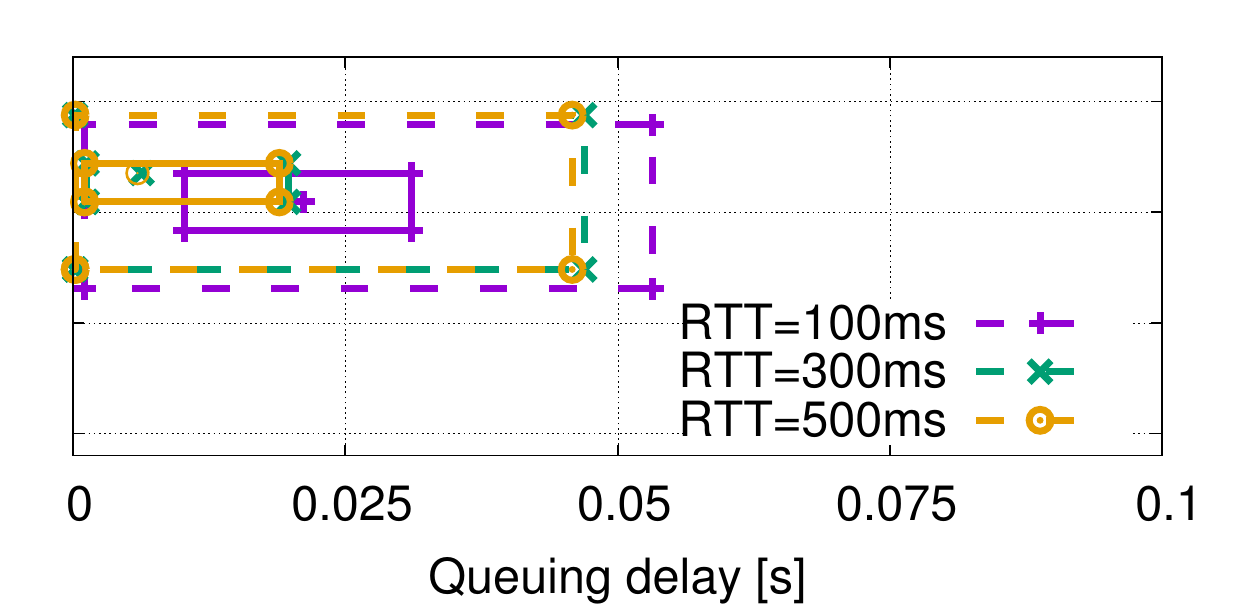}}
        \caption{Average goodput and queuing delay for the CBR traffic.}
        \label{fig-traf-mix-cbr}
    \end{center}
\end{figure*}

The performance for CBR traffic is shown in \figurename~\ref{fig-traf-mix-cbr}. 
 \figurename~\ref{fig-traf-mix-cbr-ex} explains how to interpret \figurename~\ref{subfig-traf-mix-cbr-droptail},~\ref{subfig-traf-mix-cbr-pie} and~\ref{subfig-traf-mix-cbr-madpie}. We present the average cumulative goodput as a function of the queuing delay, as advised in~\cite{rfc_aqm_charac}. 



The results with DT, shown in \figurename~\ref{subfig-traf-mix-cbr-droptail}, illustrate that queuing delay can be very high, impacting latency-sensitive applications (the higher percentiles for the queuing delay when the RTT is $500$\,ms are not shown as they do not fit in the current scale). The goodput may sometimes be over $87$\,kps as delayed packets at the bottleneck queue may arrive in bursts at the receiver. The comparison of the results of PIE (in \figurename~\ref{subfig-traf-mix-cbr-pie}) and MADPIE (in \figurename~\ref{subfig-traf-mix-cbr-madpie}) confirms that when the RTT is $100$\,ms, MADPIE does not differ much from PIE. When the RTT increases, the deterministic drops induced by MADPIE allow a reduction in the experienced queuing delay of $5$\,ms for the $75^{th}$ percentile and of $30$\,ms for the $95^{th}$ percentile, without noticeable impact on the goodput.

\subsection{Small-file downloads}
\label{subsec-eval-small-files}

\begin{figure*}[!ht]
    \begin{center}
	\subfloat[$RTT=100$\,ms. \label{subfig-traf-mix-sf-r100}]{\includegraphics[width=0.32\linewidth]{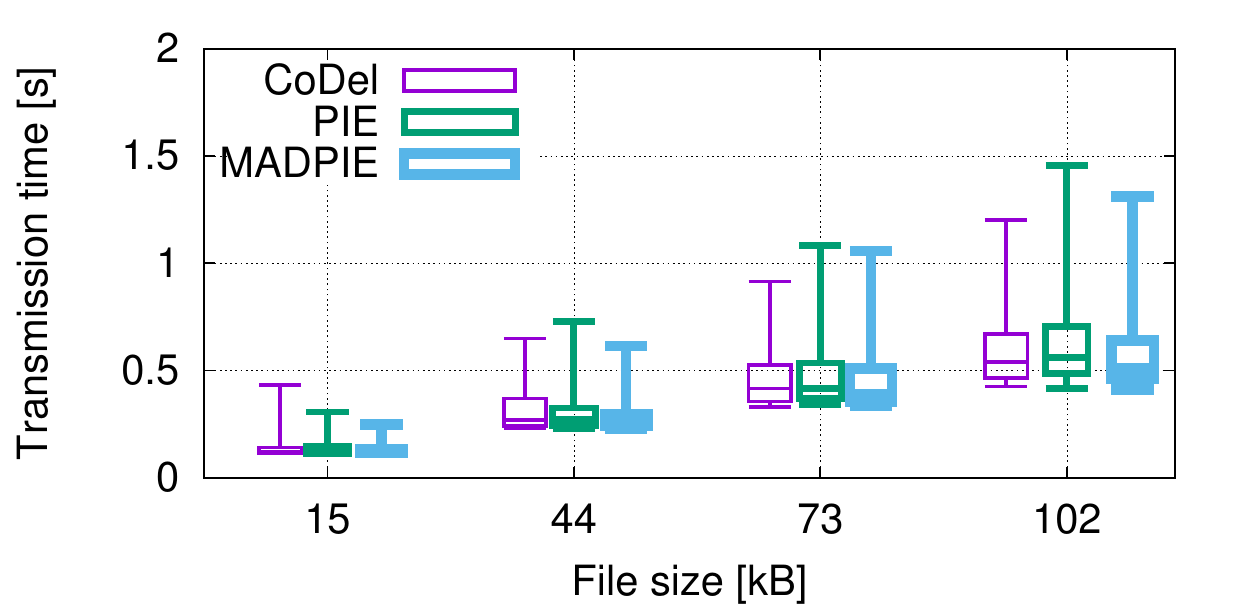}}
	\subfloat[$RTT=300$\,ms. \label{subfig-traf-mix-sf-r300}]{\includegraphics[width=0.32\linewidth]{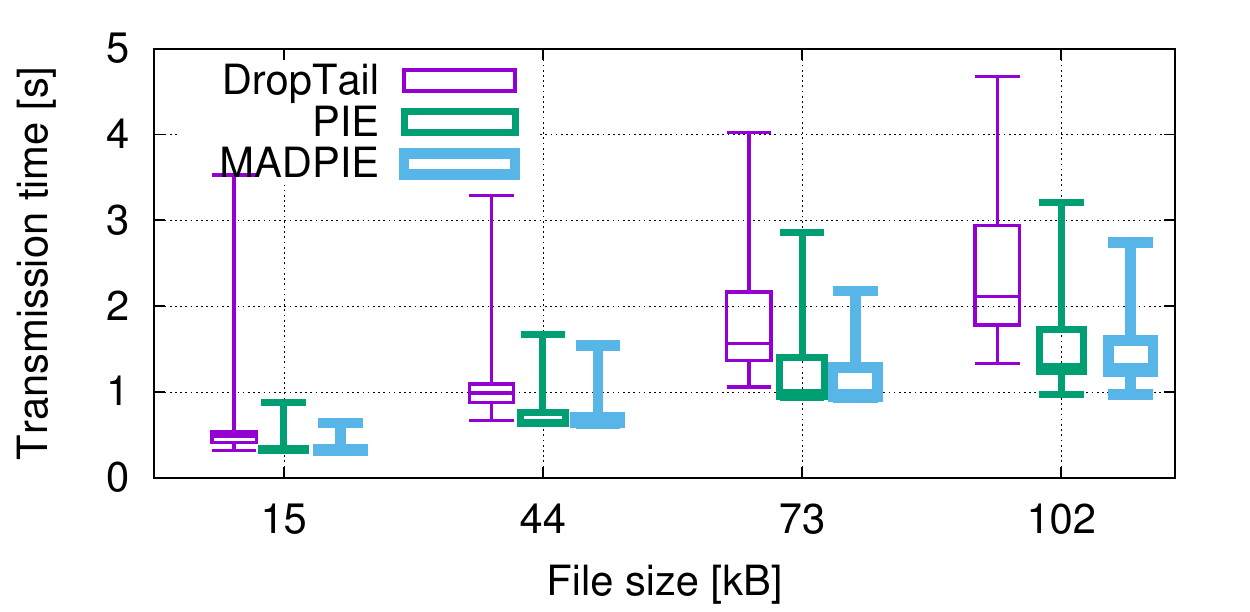}}
	\subfloat[$RTT=500$\,ms. \label{subfig-traf-mix-sf-r500}]{\includegraphics[width=0.32\linewidth]{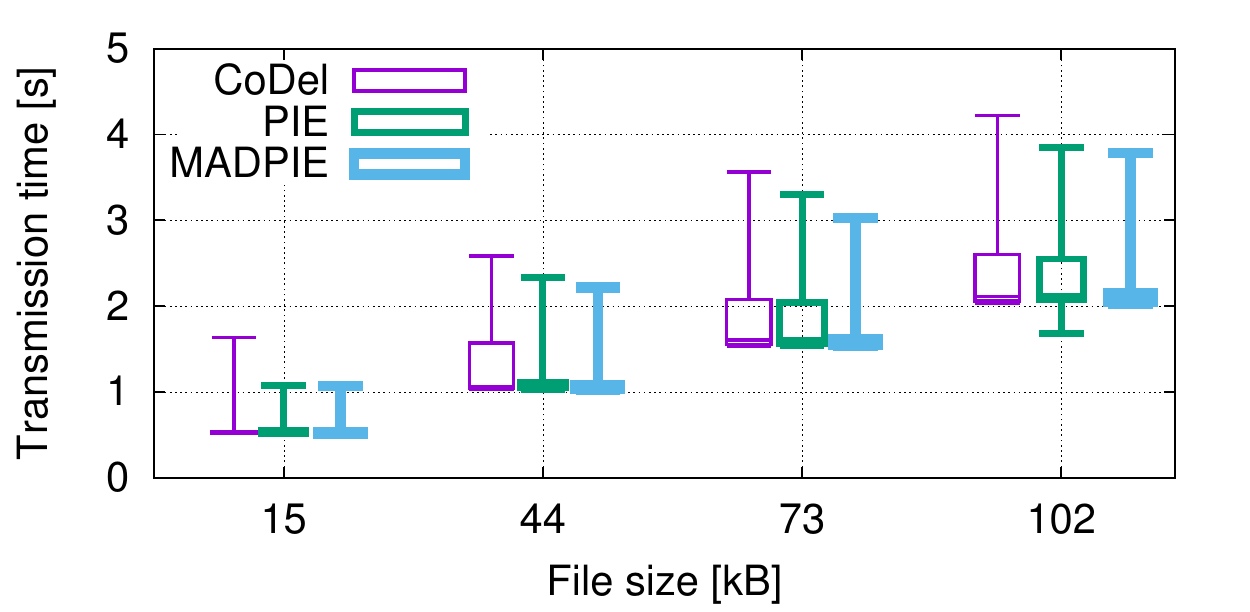}}
        \caption{Small file download time.}
        \label{fig-traf-mix-sf}
    \end{center}
\end{figure*}

We represent in \figurename~\ref{fig-traf-mix-sf} the download time of files of various sizes, with and without AQM schemes; the boxplots show the $5^{th}$, $25^{th}$, $75^{th}$ and $95^{th}$ percentiles; the line in the middle of the box is the median. With DT, the download time is higher than with any of the two AQMs for every  file size and RTT considered. Comparison of the results with MADPIE and PIE shows that MADPIE reduces worst-case transmission times. For example, with MADPIE as compared with PIE, (1) the $95^{th}$ percentile of the download time for $73$\,Kb is reduced by $\approx 700$\,ms when $RTT=300$\,ms; (2) the $75^{th}$ percentile of the download time for $102$\,Kb is reduced by $\approx 500$\,ms when $RTT=500$\,ms. 
This can be explained by the fact that, with MADPIE, the few packets that compose a short file transfer have a lower probability of experiencing high queuing delays, and of arriving at the queue when the random-drop probability is high (hence suffering losses in a burst).


\subsection{Bulk flows}
\label{subsec-eval-bulk-flows}

\begin{figure*}[!ht]
    \begin{center}
	\subfloat[$RTT=100$\,ms. \label{subfig-traf-mix-ftp-r100}]{\includegraphics[width=0.32\linewidth]{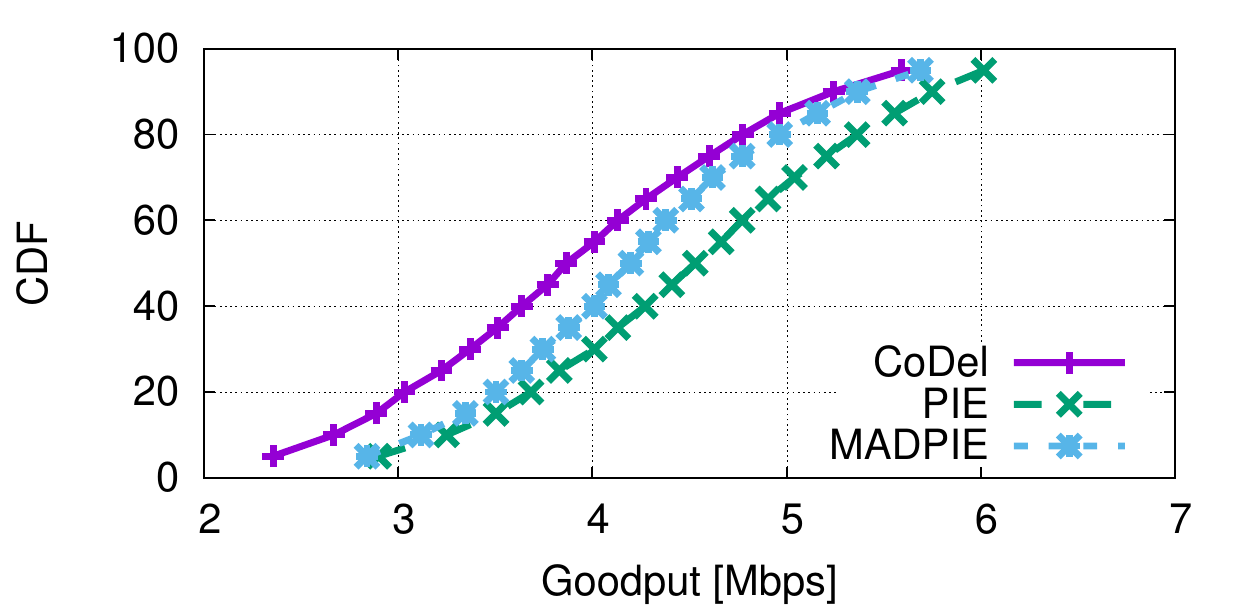}}
	\subfloat[$RTT=300$\,ms. \label{subfig-traf-mix-ftp-r300}]{\includegraphics[width=0.32\linewidth]{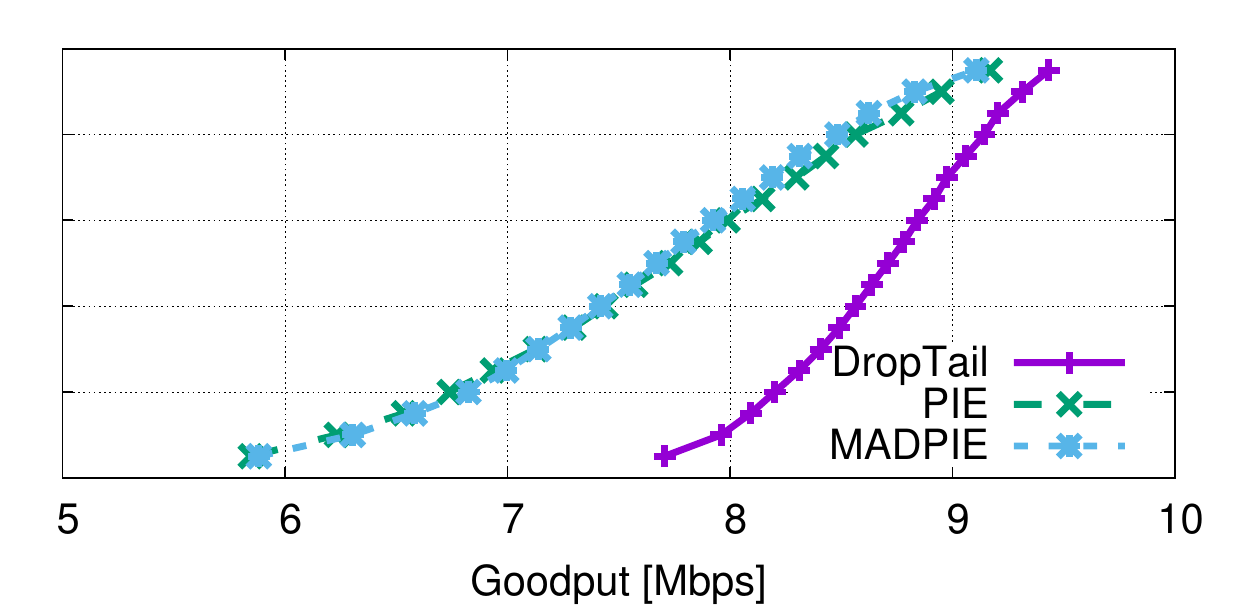}}
	\subfloat[$RTT=500$\,ms. \label{subfig-traf-mix-ftp-r500}]{\includegraphics[width=0.32\linewidth]{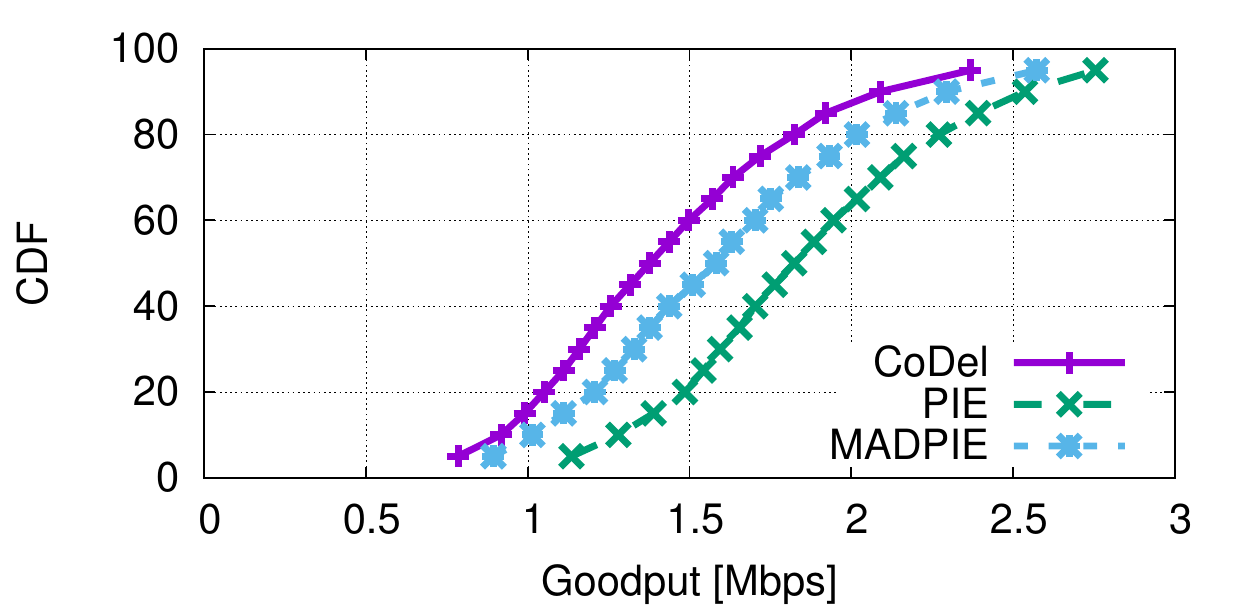}}
        \caption{Goodput of the bulk flows.}
        \label{fig-traf-mix-ftp}
    \end{center}
\end{figure*}

\figurename~\ref{fig-traf-mix-ftp} shows the CDF of the goodput for the bulk flows. With DT, the impact of the RTT can hardly be noticed. Without AQM at $R_1$, the bottleneck utilization is higher than with any of PIE or MADPIE. With the latter, when the RTT is large the gain for latency sensitive applications  comes at the expense of a small degradation in goodput for bulk flows. 


\section{Traffic mix and RTT mix}
\label{sec-rtt-mix}
Previous section focused on highlighting the difference between PIE and MADPIE; for the sake of clarity, we preferred showing results only with PIE and MADPIE. Since this section consists more of a performance analysis, rather than of a behaviour analysis, we also show results with CoDel, to compare the benefits given by introducing any of the considered AQM. We consider flows that do not face the same RTT to assess the benefits of using a scheme that is less sensitive to RTT. We do not present the results with DropTail, since \S~\ref{sec-eval} showed that it results in poor performance for latency sensitive applications when the RTT increases.
%
%
\subsection{Traffic and topology}
\label{subsec-rtt-mix-traffic-topo}
\begin{figure}[!ht]
\centering
\begin{tikzpicture}[>=stealth',shorten >=1pt,auto,node distance=2.35 cm, scale = 1.25, transform shape]
	\node[scale=0.55, color=blue]         (S) {};
	
	\node[state, scale=0.55]              (A) [right of=S,xshift=1cm] {$R_{1}$};
	\node[scale=0.55]                     (I) [below of=A, xshift=0.5cm, yshift=1.7cm] {};
        \node[state, scale=0.55]              (B) [right of=A,xshift=1.5cm] {$R_{2}$};
	
	\node[scale=0.55, color=green]        (H) [above of=S,yshift=-1.5cm] {$snd_{FTP-100}$};
	\node[scale=0.55, color=blue]         (J) [above of=S,yshift=-1cm] {$snd_{SF-100}$};
	\node[scale=0.55, color=magenta]      (G) [above of=S,yshift=-0.5cm] {$snd_{CBR-100}$};
	\node[scale=0.55, color=green]        (K) [below of=S,yshift=0.5cm] {$snd_{FTP-500}$};
	\node[scale=0.55, color=blue]         (L) [below of=S,yshift=1cm] {$snd_{SF-500}$};
	\node[scale=0.55, color=magenta]      (M) [below of=S,yshift=1.5cm] {$snd_{CBR-500}$};
	
	\node[scale=0.55] (Z1) [right of=H, xshift=-1.5cm] {};
	\node[scale=0.55] (Z2) [right of=J, xshift=-1.5cm] {};
	\node[scale=0.55] (Z3) [right of=G, xshift=-1.5cm] {};
	\node[scale=0.55] (Z4) [right of=K, xshift=-1.5cm] {};
	\node[scale=0.55] (Z5) [right of=L, xshift=-1.5cm] {};
	\node[scale=0.55] (Z6) [right of=M, xshift=-1.5cm] {};

	\node[scale=0.55, color=blue]         (D) [right of=B,xshift=1cm] {};
	\node[scale=0.55, color=blue]         (N) [above of=D,yshift=-1cm] {$dest_{SF-100}$};
	\node[scale=0.55, color=magenta]      (E) [above of=D,yshift=-0.5cm] {$dest_{CBR-100}$};
	\node[scale=0.55, color=green]        (F) [above of=D,yshift=-1.5cm] {$dest_{FTP-100}$};
	\node[scale=0.55, color=blue]         (O) [below of=D,yshift=1cm] {$dest_{SF-500}$};
	\node[scale=0.55, color=magenta]      (P) [below of=D,yshift=1.5cm] {$dest_{CBR-500}$};
	\node[scale=0.55, color=green]        (Q) [below of=D,yshift=0.5cm] {$dest_{FTP-500}$};
	
	\node[scale=0.55] (Z7) [left of=N, xshift=1.5cm] {};
	\node[scale=0.55] (Z8) [left of=E, xshift=1.5cm] {};
	\node[scale=0.55] (Z9) [left of=F, xshift=1.5cm] {};
	\node[scale=0.55] (Z10) [left of=O, xshift=1.5cm] {};
	\node[scale=0.55] (Z11) [left of=P, xshift=1.5cm] {};
	\node[scale=0.55] (Z12) [left of=Q, xshift=1.5cm] {};
%
	\path (Z2)  edge [<->, color=black] node[scale=0.5]{\small{$OWD=1$\,ms}} (A)
		    (Z3)  edge [ <->, color=black] node[scale=0.5]{\small{$100$\,Mbps}} (A)
		    (Z1)  edge [ <->, color=black] node[scale=0.5]{} (A)
		    (A)  edge [ <->, color=black] node[scale=0.5]{\small{$100$\,Mbps}} (Z5)
		    (A)  edge [ <->, color=black] node[scale=0.5]{} (Z6)
		    (A)  edge [ <->, color=black] node[scale=0.5]{\small{$OWD=201$\,ms}} (Z4)

		    (A)  edge [ <->, color=red] node[scale=0.5]{\small{$10$\,Mbps-$OWD=48$\,ms}} (B)	
		    (B)  edge [ <->, color=black] node[scale=0.5]{\small{$OWD=1$\,ms}} (Z7)
		    (B)  edge [ <->, color=black] node[scale=0.5]{\small{$100$\,Mbps}} (Z8)
		    (B)  edge [ <->, color=black] node[scale=0.5]{} (Z9)
		    (Z10)  edge [ <->, color=black] node[scale=0.5]{\small{$100$\,Mbps}} (B)
		    (B)  edge [ <->, color=black] node[scale=0.5]{} (Z11)
		    (Z12)  edge [ <->, color=black] node[scale=0.5]{\small{$OWD=1$\,ms}} (B);
\end{tikzpicture}
\caption{Topology and traffic mix used to evaluate the RTT sensitivity}
\label{fig-topology-traf-rtt-mix}
\end{figure}
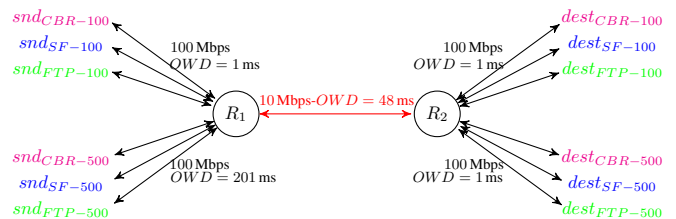
Figure~\ref{fig-topology-traf-rtt-mix} presents the topology used in this section. The traffic generated, the parameterization of PIE and MADPIE, and the methodology (number of runs, duration of each run) have the same characteristics as in \S~\ref{sec-eval}, the only difference being the number of flows that is the following: $N_{CBR-100}=N_{CBR-500}=4$, $N_{SF-100}=N_{SF-500}=20$ and $N_{FTP-500}=N_{FTP-500}=2$, where $N_{X-100}$ represents the number of flows for the application of type $X$ on the path with $100$\,ms of RTT. As the RTT of the paths are not the same, the queue size at $R_1$ is set to the BDP of the higher RTT.
\subsection{CBR traffic - between $snd_{SF-X}$ and $dest_{SF-X}$}
\label{subsec-rtt-mix-cbr}
\begin{figure}[!ht]
    \begin{center}
	\includegraphics[width=\linewidth]{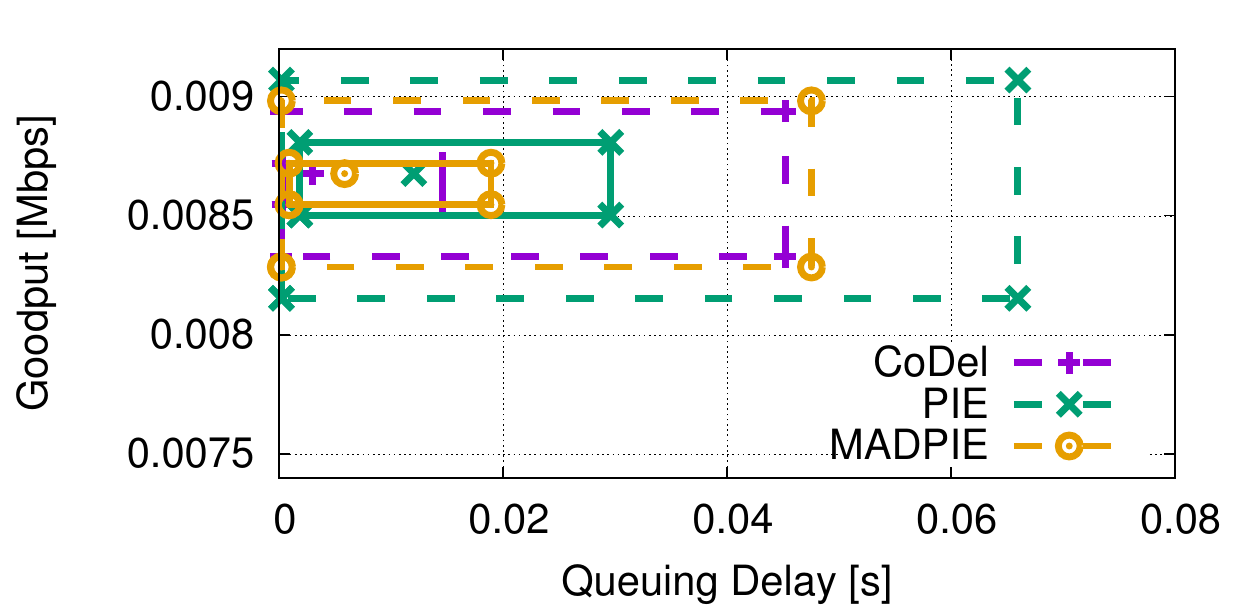}
        \caption{Goodput and queuing delay - CBR traffic for which $RTT=100$\,ms}
        \label{fig-traf-rtt-mix-cbr}
    \end{center}
\end{figure}

The results for the CBR traffic are shown in Figure~\ref{fig-traf-rtt-mix-cbr} and we use the same representation as in Figure~\ref{fig-traf-mix-cbr-ex}. Due to the presence of flows that experience an RTT of $500$\,ms, the flows with an RTT of $100$\,ms face a queuing delay that momentarily rises above $60$\,ms with PIE (\textit{i.e.} $55$\,\% of the one way delay). Since the default \textit{target delay} of CoDel is $5$\,ms, the allowed queuing delay is lower than with PIE and MADPIE, for which the \textit{target delay} is set to $20$\,ms. By default, CoDel would maintain a lower queuing delay than PIE. Also, because CoDel uses deterministic drops, the queuing delay can not rise much higher before the first drops are applied. These results show that MADPIE takes the best of the two schemes: with MADPIE, the median queuing delay is lower than with PIE and the queuing delay is kept under control, and the higher percentiles of the queuing delay are close to those with CoDel. While with PIE, the queuing delay momentarily increases above $60$\,ms (\textit{i.e.} $55$\,\% of the RTT), with MADPIE, the introduction of the flows that experience an RTT of $500$\,ms has less impact: the queuing delay momentarily rises above $40$\,ms, that is $45$\,\% of the RTT. MADPIE provides a latency reduction of $\approx 13$\,\% for the $75^{th}$ percentiles and of $\approx 21$\,\% for the $95^{th}$ percentiles and the performance is close to that of CoDel.
 
\subsection{Small files download - between $snd_{SF-X}$ and $dest_{SF-X}$}
\label{subsec-rtt-mix-small-files}

\begin{figure}[!ht]
    \begin{center}
	\subfloat[$RTT=100$\,ms \label{subfig-traf-rtt-mix-sf-r100}]{\includegraphics[width=0.95\linewidth]{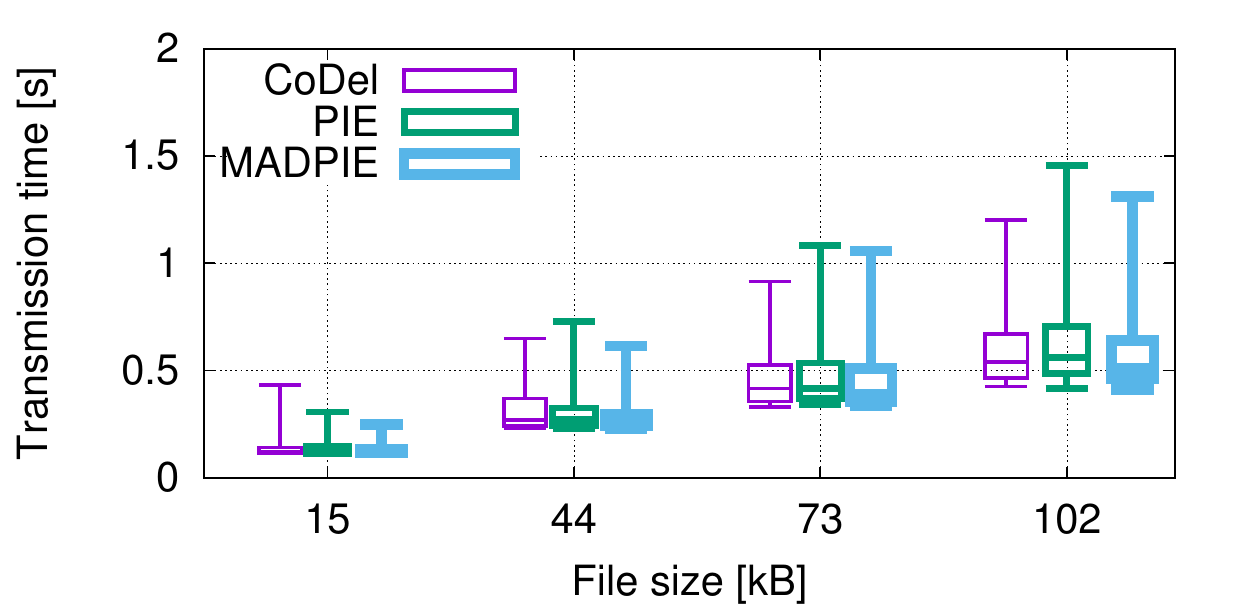}} \\
	\subfloat[$RTT=500$\,ms \label{subfig-traf-rtt-mix-sf-r500}]{\includegraphics[width=0.95\linewidth]{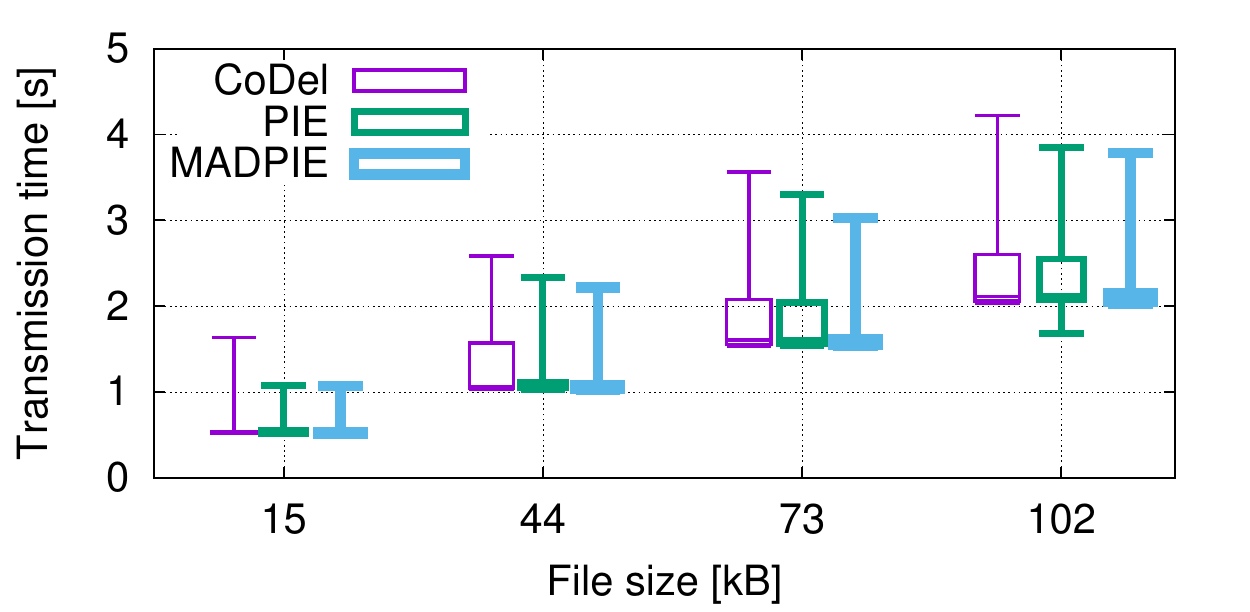}}
        \caption{Small file download time}
        \label{fig-traf-rtt-mix-sf}
    \end{center}
\end{figure}

We show in Figure~\ref{fig-traf-rtt-mix-sf} the download time of small files for the flows that experience an RTT of $100$\,ms (Figure~\ref{subfig-traf-rtt-mix-sf-r100}) and of $500$\,ms (Figure~\ref{subfig-traf-rtt-mix-sf-r500}).

For small file downloads over the path with a base RTT of $100$\,ms, the download time is lower than in~\S~\ref{subsec-eval-small-files} as there is much less concurrent traffic: in this section, there are $4$ bulk flows as opposed to $10$ in \S~\ref{sec-eval}. The $5^{th}$, $25^{th}$, $50^{th}$ and $75^{th}$ percentiles are quite the same whether the AQM is PIE, MADPIE or CoDel; the $75^{th}$ percentile is slightly lower with MADPIE. The $95^{th}$ percentile is always lower with MADPIE. When the file size is larger than $73$\,kB, the $95^{th}$ percentile is slightly lower with CoDel or MADPIE than with PIE, probably thanks to the fact that if the AQM drops a packet, the queuing delay experienced by the retransmission would be lower. 
For small file downloads that experience a base RTT of $500$\,ms, the performance is quite the same with CoDel and PIE: with PIE, the bottleneck utilisation is higher (see \S~\ref{subsec-rtt-mix-bulk-flows} for more details) and with CoDel the queuing delay is lower (see \S~\ref{subsec-rtt-mix-cbr} for more details). MADPIE takes the best of both schemes and thus, provides lower download times for small files. 
\subsection{Bulk flows - between $snd_{FTP-X}$ and $dest_{FTP-X}$}
\label{subsec-rtt-mix-bulk-flows}
\begin{figure}[!ht]
    \begin{center}
	\subfloat[$RTT=100$\,ms \label{subfig-traf-rtt-mix-ftp-r100}]{\includegraphics[width=0.95\linewidth]{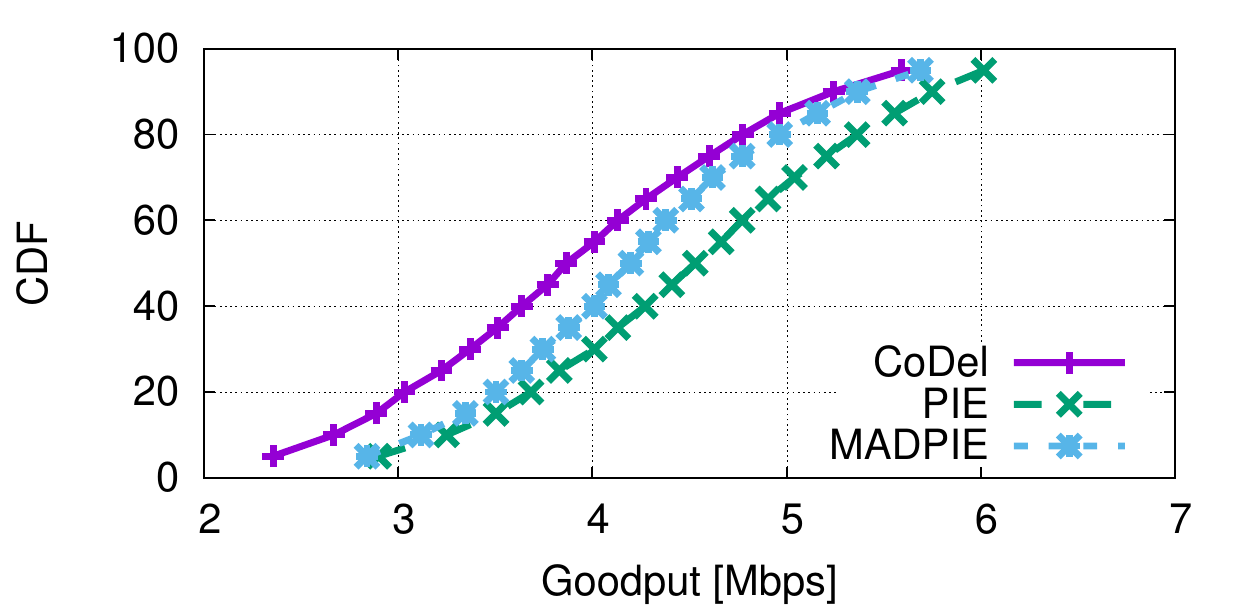}} \\
	\subfloat[$RTT=500$\,ms \label{subfig-traf-rtt-mix-ftp-r500}]{\includegraphics[width=0.95\linewidth]{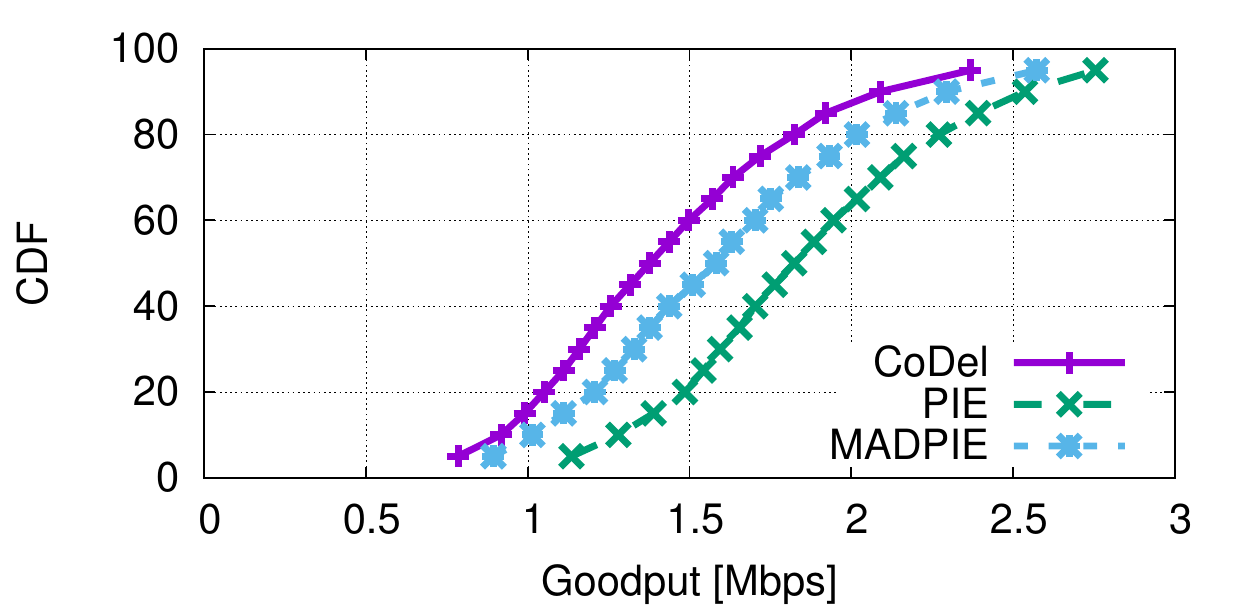}}
        \caption{Goodput of the bulk flows}
        \label{fig-traf-rtt-mix-ftp}
    \end{center}
\end{figure}

We show in Figure~\ref{fig-traf-rtt-mix-ftp} the goodput of bulk transfers for flows that experience an RTT of $100$\,ms (Figure~\ref{subfig-traf-rtt-mix-ftp-r100}) and of $500$\,ms (Figure~\ref{subfig-traf-rtt-mix-ftp-r500}). CoDel shows a lower goodput than PIE, which is due to its lower \textit{target delay}: PIE allows more buffering. As seen in \S~\ref{subsec-eval-bulk-flows}, MADPIE slightly reduces the bottleneck utilization for the bulk flows that experience an RTT of $100$\,ms. The same happens for the flows that experience an RTT of $500$\,ms. With MADPIE, the resulting goodput is a trade-off between CoDel and PIE.

\subsection{Discussion}
\label{subsec-rtt-mix-discussion}
%
%
%
The benefits of using MADPIE instead of CoDel may not be clear, however we advice to deploy MADPIE instead of CoDel. Indeed, in~\cite{jarvinen-lcn-2014}, CoDel has been shown to have auto-tuning issues and its performance are sensitive to the load of traffic. Also, $5$\,ms of maximum allowed queuing delay can be damaging for bottlenecks of $2$\,Mbps~\cite{2015-kuhn-eucnc} and its interval value is based on the assumption that the RTT is $100$\,ms~\cite{ietf-codel}, which is not the case for rural broadband networks. 
On the contrary, our algorithm clearly improves the performance of PIE when the RTT is higher than $300$\,ms for various types of traffic and does not affect the performance of PIE for lower RTTs. The deployment issues of CoDel mentioned earlier in this section are solved with PIE as: (1) it is less sensitive to traffic loads~\cite{jarvinen-lcn-2014}; (2) with $20$\,ms of targeted queuing delay, we expect less issues with low capacity bottlenecks; (3) it does not make assumptions on the RTT. 


\section{Conclusion}
\label{sec-conclusions}

In this paper, we have proposed  MADPIE, a simple change to the PIE algorithm that makes it less dependent on path RTTs in lightly-multiplexed scenarios. MADPIE extends PIE by adding, on top of the random drops, a deterministic drop policy loosely based on CoDel's. The proportion of deterministic drops increases when the RTT increases. 
MADPIE can both keep the same target delay as PIE and reduce the maximum queuing delay, making the goodput of bulk flows close to the one achieved with PIE, guaranteeing lower queuing delay for VoIP-like traffic and reducing the download time of small files.

We do not claim that our proposal is the only, or best, way of tuning or adapting PIE. However, it is a very simple addition to PIE's code (a handful of lines, in our ns-2 implementation) that can complement specific parameter tunings, and its impact on the performance of PIE seems negligible when RTTs are not large (i.e, outside the operating conditions for which it has been conceived).




 


\section*{Acknowledgement}
This work was partially funded by the European Community under its Seventh Framework Programme through the Reducing Internet Transport Latency (RITE) project (ICT-317700).

\bibliographystyle{IEEEtran}
\bibliography{hybrid-pie-codel}

\end{document}